\documentclass[prl,twocolumn,showpacs,preprintnumbers,amsmath,amssymb,floatfix]{revtex4}

\usepackage{graphicx}
\usepackage{dcolumn}
\usepackage{bm}

\usepackage[normalem]{ulem}  
\usepackage[dvips]{color}

\begin{document}


\title{Isospin effects and the density dependence of the nuclear symmetry energy}

\author{S.R.\ Souza$^{1,2}$}
\author{M.B.\ Tsang$^3$}
\author{B.V.\ Carlson$^{4}$}
\author{R.\ Donangelo$^{1,5}$}
\author{W.G.\ Lynch$^3$}
\author{A.W.\ Steiner$^3$}
\affiliation{$^1$Instituto de F\'\i sica, Universidade Federal do Rio de Janeiro,
Cidade Universit\'aria, \\CP 68528, 21941-972, Rio de Janeiro, Brazil}
\affiliation{$^2$Instituto de F\'\i sica, Universidade Federal do Rio Grande do Sul\\
Av. Bento Gon\c calves 9500, CP 15051, 91501-970, Porto Alegre, Brazil}
\affiliation{$^3$ Joint Institute for Nuclear Astrophysics, National
Superconducting Cyclotron Laboratory, and the Department of Physics
and Astronomy, Michigan State University,
East Lansing, MI 48824, USA}
\affiliation{$^4$Departamento de F\'\i sica, 
Instituto Tecnol\'ogico de Aeron\'autica - CTA, 12228-900\\
S\~ao Jos\'e dos Campos, Brazil}
\affiliation{$^5$Instituto de F\'\i sica, 
Facultad de Ingenier\'\i a, Universidad de la Rep\'ublica,\\
Julio Herrera y Reissig 565, 11.300 Montevideo, Uruguay}

\date{\today}

\begin{abstract}
The density dependence of the nuclear symmetry energy is inspected using the Statistical Multifragmentation Model
with Skyrme effective interactions.
The model consistently considers the expansion of the fragments' volumes at
finite temperature at the freeze-out stage.
By selecting parameterizations of the Skyrme force that lead to very different equations of state for the
symmetry energy, we investigate the sensitivity of different observables to the properties of the effective forces.
Our results suggest that, in spite of being sensitive to the thermal dilation of the fragments' volumes, it
is difficult to distinguish among the Skyrme forces from the isoscaling analysis.
On the other hand, the isotopic distribution of the emitted fragments turns out to be very sensitive to the force
employed in the calculation.
\end{abstract}

\maketitle

Investigations on the density dependence of the symmetry energy in the nuclear multifragmentation process have been stimulated by the
discovery \cite{isoscaling1,isoscaling2} that the ratio $R_{21}$ between the experimental yields $Y(A,Z)$ of a
fragment of mass and atomic numbers $A$ and $Z$, respectively, produced in similar reactions with different isospin
compositions, henceforth labeled `1' and `2',
follows a scaling law
\cite{isoscalingAMD2003,RadutaIsoSym1,RadutaIsoSym2,isoMassFormula2008,
isoscalingIndraGSI2005,symEnergyShatty2007,BotvinaSurf2006,isoNatowitz2007_2,isoSouliotis2007,
IsospinSymmetry,isoSymmetryBotvina2006,isocc,isoscaling3,isoscWolfgangBotvina1}.
This isoscaling law provides parameters $\alpha$ and $\beta$ that are determined from the property

\begin{equation}
R_{21}\equiv \frac{Y_2(A,Z)}{Y_1(A,Z)}=C\exp(\alpha N + \beta Z)\;
\label{eq:iso}
\end{equation}

\noindent
where $C$ is a normalization constant \cite{isoscaling1}.

Studies of the density dependence of the symmetry energy have used Eq.\ (\ref{eq:iso}) to: ({\it i}) probe 
the dependence of statistical models on the symmetry energy
\cite{isoscalingIndraGSI2005,symEnergyShatty2007,BotvinaSurf2006,isoSymmetryBotvina2006} and ({\it ii}) to probe
the isospin composition of the region emitting the fragments \cite{isoscalingAMD2003,TanisoDiff,TanIsoEOS}.
In this paper, we are concerned with the former issue.
The motivation for statistical model studies arises from the relationship between the isoscaling parameter $\alpha$ and the
nuclear symmetry energy, whose leading term at low temperatures was shown to be \cite{isoscaling3,isoscWolfgangBotvina1}

\begin{equation}
\alpha\approx 4C_{\rm sym} \left[(Z_1/A_1)^2-(Z_2/A_2)^2\right]/T\;,
\label{eq:alphaSymEner}
\end{equation}

\noindent
where $C_{\rm sym}$ denotes the symmetry energy coefficient \cite{isoMassFormula2008},
$T$ is the temperature of the system,
$Z_i$ and $A_i$, $i=1,2$, correspond, respectively, to the atomic and mass numbers of the decaying source.

To be useful, reactions `1' and `2' should be chosen to produce systems at approximately the same temperature and density.
The ratio in Eq.\ (\ref{eq:iso}) involves yields of the same isotope; distorting effects associated with the
deexcitation of the primordial hot fragments may be similar in the two reactions and lead to an approximate cancellation in the
ratio.
Theoretical calculations support this assumption for primary distributions calculated from equilibrium statistical models 
\cite{isoscaling3,isocc,isoSymmetryBotvina2006}.
Therefore, measurements of the isoscaling parameters, through fits based on
Eq.\ (\ref{eq:iso}), may provide valuable information on the symmetry energy.

This assumption has extensively been exploited in many works \cite{isoscalingIndraGSI2005,isoSymmetryBotvina2006,symEnergyShatty2007,
BotvinaSurf2006,isoNatowitz2007_2,isoSouliotis2007,IsospinSymmetry} which employed the Statistical Multifragmentation Model
(SMM) \cite{Bondorf1995}.
In order to reproduce the measured $\alpha$ parameter, $C_{\rm sym}$ has been appreciably reduced, compared
with the values usually adopted in this model.
The main conclusion of those works \cite{isoscalingIndraGSI2005,isoSymmetryBotvina2006,symEnergyShatty2007,
BotvinaSurf2006,isoNatowitz2007_2,isoSouliotis2007,IsospinSymmetry}
is that there seems to be an important decrease of the symmetry energy at low densities.

Although this result is reasonable on physical grounds, other studies \cite{RadutaIsoSym1,RadutaIsoSym2,isoMassFormula2008}
also provided a sound explanation to this apparent reduction of $C_{\rm sym}$.
They suggest that surface effects associated with  the symmetry energy, not considered in Refs.\
\cite{isoscalingIndraGSI2005,isoSymmetryBotvina2006,symEnergyShatty2007,BotvinaSurf2006,isoNatowitz2007_2,isoSouliotis2007,
IsospinSymmetry}, may also lead to significant reduction of the $\alpha$ parameter.
This explanation seems to be more reasonable since the model used in
all these studies \cite{isoscalingIndraGSI2005,isoSymmetryBotvina2006,symEnergyShatty2007,
BotvinaSurf2006,isoNatowitz2007_2,isoSouliotis2007,IsospinSymmetry,RadutaIsoSym1,RadutaIsoSym2,isoMassFormula2008}
is based on binding energy formulae that evaluate the isoscaler volume, surface and Coulomb terms at the ground state (saturation)
density.
Although the volume occupied by the total system is much larger than that of the ground state source $V_0$,
the lower density values used in these statistical calculations are due to the space between the fragments.
Therefore, in this scenario, there should be no sensitivity to the density dependence of the bulk symmetry energy.

In this work we investigate this issue in a consistent approach in which the properties of the fragments
are calculated through the Thomas-Fermi approximation (TFA) at finite temperature.
This version of the SMM, named SMM-TF, was presented in Ref.\ \cite{smmtf1}.
In this way, the changes to the fragments' energies and densities at the freeze-out stage are consistently obtained
in the framework of the TFA.
Thus, the total Helmholtz free energy $F$ of a given partition mode is written in the same form as in
the Improved Statistical Multifragmentation Model (ISMM) \cite{ISMMlong}, also used in this work:

\begin{eqnarray}
\label{eq:fe}
&& F(T)=\frac{C_{\rm Coul}}{(1+\chi)^{1/3}}\frac{Z_0^2}{A_0^{1/3}}+F_{\rm trans}(T)\\
&&+\sum_{A,Z}N_{A,Z}\left[-B_{A,Z}+f^*_{A,Z}(T)-\frac{C_{\rm Coul}}{(1+\chi)^{1/3}}\frac{Z^2}{A^{1/3}}\right]\nonumber
\end{eqnarray}

\noindent
where

\begin{eqnarray}
 F_{\rm trans}=-T(M-1)\log\left(V_f/\lambda_T^3\right)+T\log\left(g_0A_0^{3/2}\right)&&\\
 -T\sum_{A,Z}N_{A,Z}\left[\log\left(g_{A,Z}A^{3/2}\right)-\frac{1}{N_{A,Z}}\log(N_{A,Z}!)\right]\;.&&\nonumber
\label{eq:fekin}
\end{eqnarray}

\noindent
In the above Eqs., $C_{\rm Coul}$ denotes the Coulomb coefficient of the mass formula \cite{ISMMlong},
$A_0$ and $Z_0$ are the mass and atomic numbers of the decaying source, $B_{A,Z}$ represents the binding energy of the
fragment, $N_{A,Z}$ stands for its multiplicity and $M=\sum_{A,Z}N_{A,Z}$.
The freeze-out volume $V_\chi=V_0(1+\chi)$ is kept fixed ($\chi=2$) for all the values of the source's excitation energy $E^*$.
The factor $M-1$, rather than $M$, as well as $T\log\left(g_0A_0^{3/2}\right)$, in the translational contribution to the free
energy, arise from the subtraction of the center of
mass motion from the partition function of the total system.
The spin degeneracy factor is denoted by $g_{A,Z}$, $\lambda_T=\sqrt{2\pi\hbar^2/m_nT}$ corresponds to the thermal wave-length, and
$m_n$ is the nucleon mass.
As in the ISMM, the free volume reads

\begin{equation}
V_f(T)=V_0(1+\chi)-\sum_{A,Z}N_{A,Z}V_{A,Z}(T)\;,
\label{eq:vfree}
\end{equation}

\noindent
where $V_{A,Z}(T)$ is the volume occupied by each fragment.
However, in the SMM-TF, $V_{A,Z}(T)$ is given by the TFA \cite{smmtf1} and therefore it differs from the ground state
value used in the ISMM.
The internal Helmholtz free energy of the fragment $f^*_{A,Z}$ in the SMM-TF model is also given by this microscopic approach.
These are the only two differences between the SMM-TF and the ISMM, as discussed in Ref. \cite{smmtf1}.
All the other ingredients are the same.

\begin{figure}[tb]
\includegraphics[width=8.2cm]{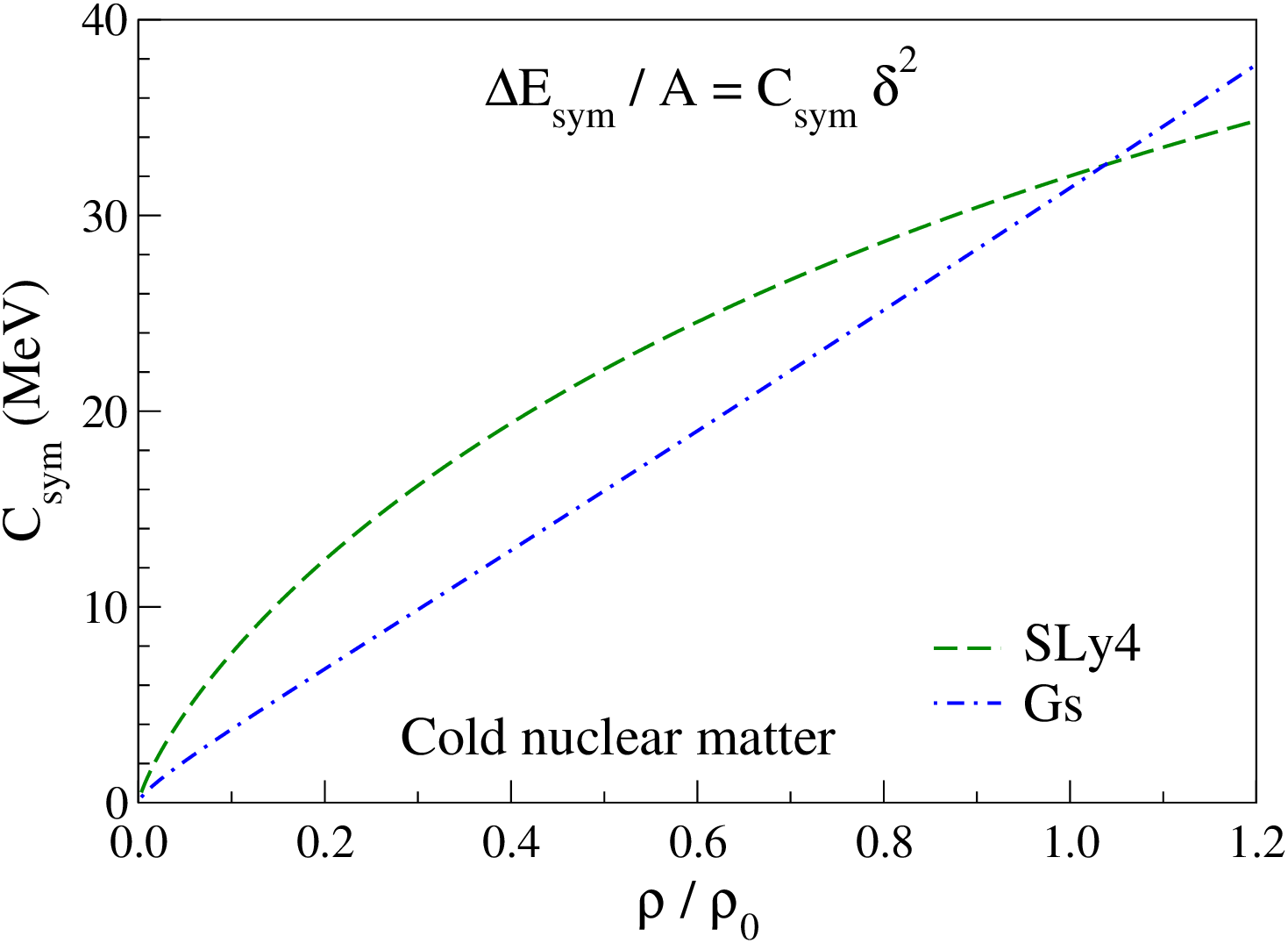}
\caption{\label{fig:nucMat} (Color online) Symmetry energy coefficient of cold nuclear matter for the SLy4 and Gs Skyrme forces
as a function of the density $\rho$.
For details, see the text.
}
\end{figure}

Owing to the plethora of Skyrme forces in the literature \cite{SkyrmeForces1,HartreeFockPawel1}, we do not perform a detailed study using
all the existing parameterizations.
Instead, we selected two of them which give distinct Equations of State (EOS's) for the symmetry energy, {\it i. e.}, the Gs
\cite{SkyrmeGs} and the SLy4 \cite{SkyrmeSLy4} forces.
Figure \ref{fig:nucMat} shows the density dependence of the symmetry energy coefficient of cold nuclear matter,
$\delta E_{\rm sym}/A=C_{\rm sym}\delta ^2$, where $\delta=(\rho_n-\rho_p)/(\rho_n+\rho_p)$ and $\rho_n$ ($\rho_p$) is the neutron
(proton) density.
One sees that, although both forces agree for densities close to the saturation value ($\rho_0$), the differences at lower densities
can be appreciably large.
The Gs and SLy4 forces provide examples of a strongly density dependent (stiff) and a weakly density dependent (soft) symmetry energy,
respectively.
Both have bulk isoscaler incompressibility moduli in the range of 230 to 250 MeV \cite{EOSBonche}.
Therefore, they are well suited to the present study.

We confine our attention to the $^{40}$Ar,$^{40}$Ca $+$ $^{58}$Ni systems, for which
measurements of the isoscaling parameter $\alpha$ have been reported recently \cite{isoSymmetryBotvina2006}.
The decaying sources considered in the calculations below correspond to 80\% of the compound systems,
as 20\% of the matter is removed in order to take the preequilibrion emission into account.
To make a connection to the work of Ref.\ \cite{isoSymmetryBotvina2006}, we adopt in the following their assumption that the $Z/A$ of the
source is the same as the original system. 

We begin our discussion by examining whether the isoscaling property should still
hold if the thermal expansion of the fragments' volumes is taken into account.
The microcanonical treatment employed in the SMM-TF does not allow one to derive analytical expressions to investigate this issue.
Furthermore, the dependence of the free volume on the species multiplicities, Eq.\ (\ref{eq:vfree}), lead to highly non-linear terms in
the Helmholtz free energy, rendering the traditional grand-canonical formulas \cite{isoMassFormula2008} invalid.
By minimizing $F$ with respect to the multiplicities, as is done in Ref.\ \cite{isoMassFormula2008}, one may nevertheless obtain
formal expressions which suggest that the isoscaling property should still be observed in this case.

\begin{figure}[t]
\includegraphics[width=7.9cm]{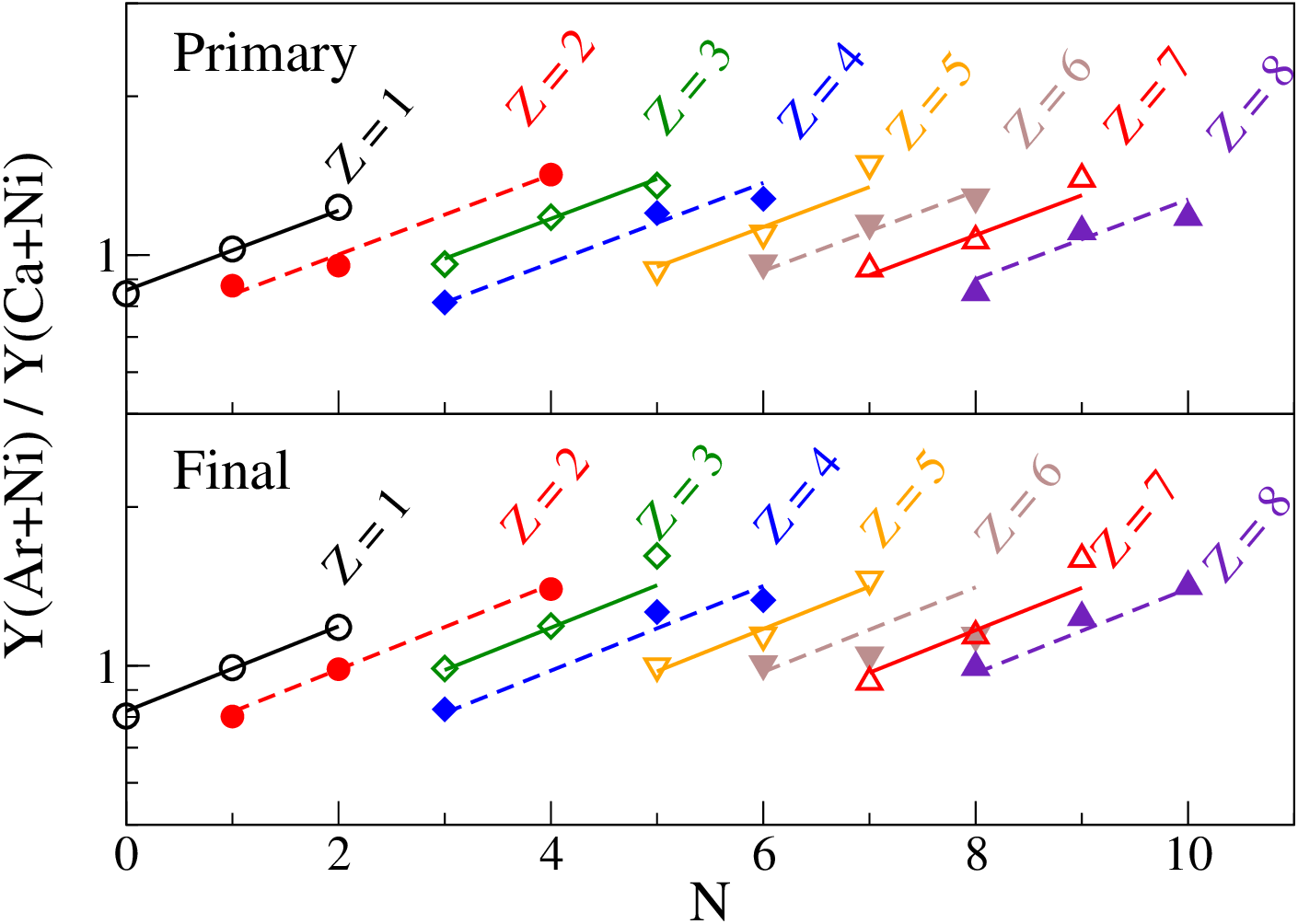}
\caption{\label{fig:isoscaling} (Color online) Isoscaling predicted by the SMM-TF using the SLy4 for $E^*/A=6.0$~MeV.
In the upper panel the yields of the hot primary fragments are used whereas the lower panel displays the results
after the deexcitation of these fragments.
The lines correspond to the best fit of the results using Eq.\ (\ref{eq:iso}).
For details, see the text.
}
\end{figure}

This is indeed found in our numerical microcanonical SMM-TF calculations, as is illustrated in Fig.\ \ref{fig:isoscaling}, which
shows $R_{21}$ for the SLy4 force and $E^*/A=6.0$~MeV.
Similar results are obtained for the Gs force and for other excitation energies.
The magnitude of the corrections due to the deexcitation of the hot primary fragments
can be estimated by comparing the upper and lower panels of Fig.\ \ref{fig:isoscaling}.
Here, we simulated the decay with a simplified Monte Carlo Weisskopf model, which includes the emission of nuclei up to oxygen.
The parameterization of the cross-section for the inverse reaction was taken from Ref.\ \cite{grandCanonicalBotvina1987}.
Following Ref.\ \cite{smmDecay}, we calculate the density of states from the entropies associated with $f^*_{A,Z}$.
This provides a consistent link between the primary stage and the deexcitation process.
In agreement with previous calculations \cite{isoscaling3,isoSymmetryBotvina2006}, our results also suggest that $\alpha$
is not strongly sensitive to the deexcitation of the primary fragments.

\begin{figure}[t]
\includegraphics[width=8.0cm]{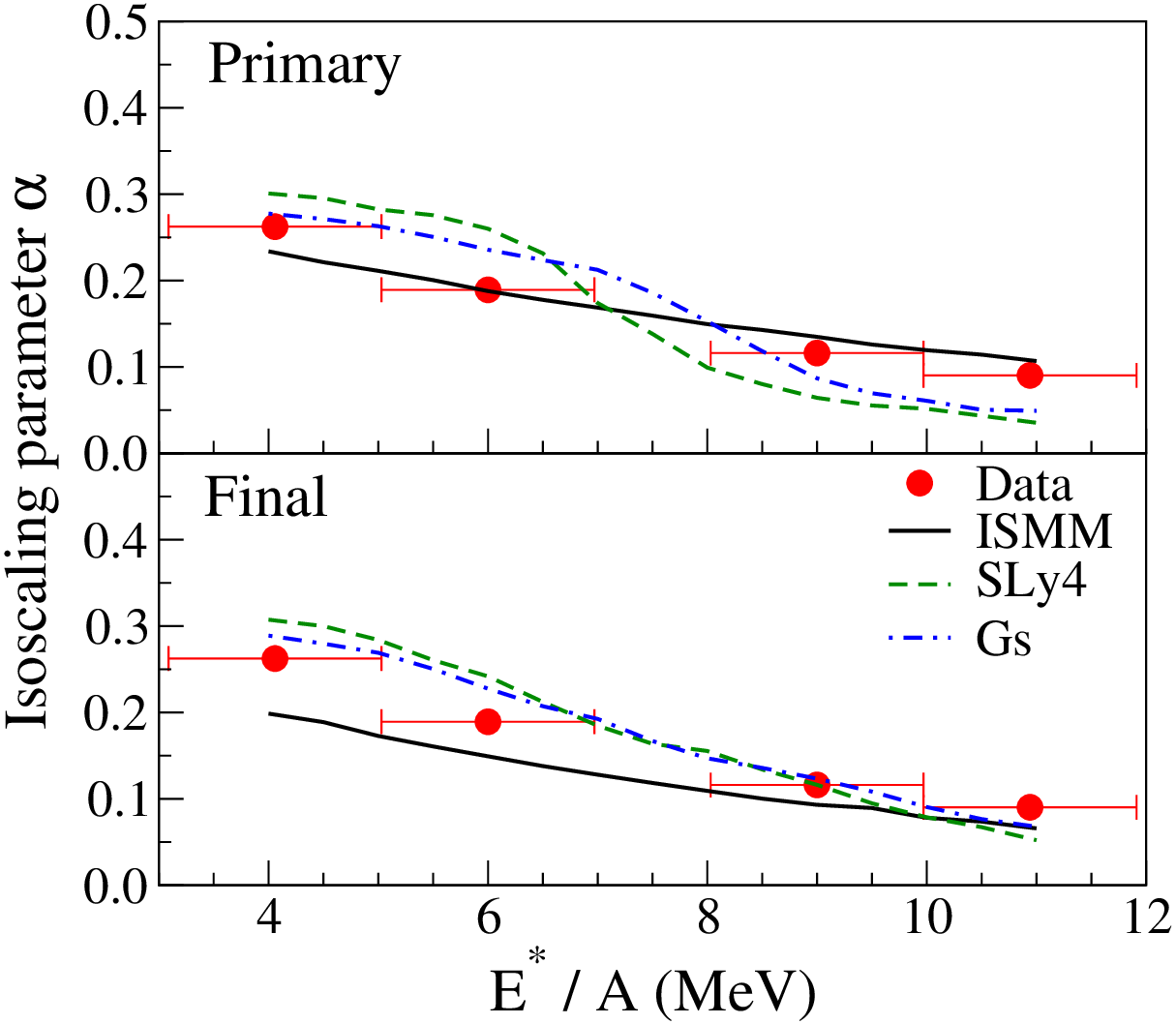}
\caption{\label{fig:alpha} (Color online) Isoscaling parameter $\alpha$ as a function of the excitation energy
of the source, for different Skyrme forces and the ISMM, before and after the deexcitation of the primary fragments.
The data are taken from Ref.\ \cite{isoSymmetryBotvina2006}.
For details, see the text.
}
\end{figure}

Figure \ref{fig:alpha} shows the comparison between the $\alpha$ values obtained in the different SMM models used in this work and
the available experimental data \cite{isoSymmetryBotvina2006}.
The results corresponding to the primary fragments (top panel of this picture) reveal that $\alpha$ is fairly sensitive to the thermal
dilatation of the fragments' volumes as the behavior of
$\alpha$ in the SMM-TF calculations is different from that given by the ISMM.
Switching from the SLy4 to the Gs force leads to small differences in $\alpha$, primarily within the
small energy range $7.0$~MeV $ < E^*/A < 8.5 $~MeV. 
This corresponds to the region where the average fragments' density is approximately $0.65\rho_0$.
For the ISMM, investigations of the connection between $\alpha$ and $C_{\rm sym}$ at finite temperatures
show that Eq.\ (\ref{eq:alphaSymEner}) can be fairly inaccurate at high temperatures,
although the main conclusion that both quantities are strongly correlated remains valid \cite{isotemp}.

It is interesting to examine whether comparisons of the SMM-TF calculations to experimental fragmentation data can clarify questions
regarding the adequacy of the model and the effective forces selected to describe the multifragment emission.
The ISMM results seem to follow the experimental trends more closely than the SMM-TF calculations.
This picture does not survive the deexcitation of the primary fragments.
Although the changes to $\alpha$ are small, the values predicted by the ISMM are systematically lowered, so that
after secondary decay it lies below the data and the SMM-TF.
The differences between the calculations with the SLy4 and the Gs forces are very small after the decay of the
primary fragments and our results suggest that it is difficult to distinguish between them from the isoscaling
analysis.

\begin{figure}[tb]
\includegraphics[width=8.0cm]{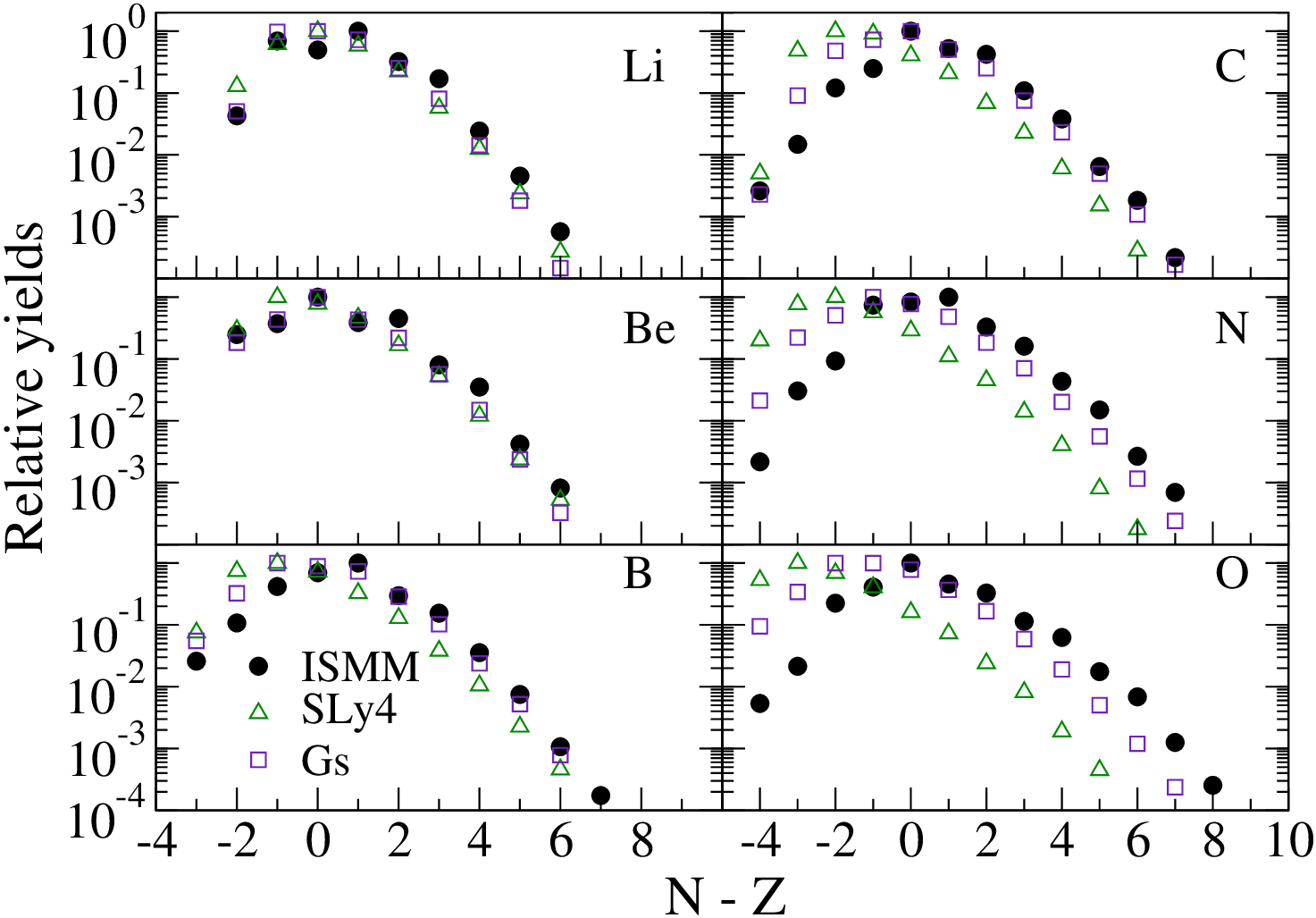}
\caption{\label{fig:iso1} (Color online) Isotopic distribution of selected primary fragments for the
Ar~$+$~Ni system at $E^*/A=8.5$~MeV.
For details see the text.
}
\end{figure}

Distinct Skyrme forces lead to different values of $V_{A,Z}(T)$, according to their properties.
This directly affects the Helmholtz free energy of the system through changes in $V_f$
and indirectly through changes in the internal free energies of the fragments.
Therefore, it suggests that the fragment composition should be very sensitive to the thermal expansion of $V_{A,Z}$.
This is indeed observed in Fig.\ \ref{fig:iso1}, which displays the isotopic distribution of
selected primary hot nuclei for the breakup of the Ar~$+$~Ni system at $E^*/A=8.5$~MeV.
It has been shown in Ref.\ \cite{smmtf1} that, at low energies, the isotopic distributions predicted by the SMM-TF
are narrower than those given by the ISMM.
However, they are very different at higher excitation energies, as shown in Fig.\ \ref{fig:iso1}, where, for this system,
the peak of the distribution shifts toward the proton rich isotopes.
The effect is enhanced for higher excitation energies and
it is more pronounced in the case of the SLy4 force than for the Gs force.
We have checked that, although these trends remain true, the differences between the Gs and the SLy4 forces are reduced after
the decay of the primary fragments whereas they remain large enough to clearly distinguish between the ISMM and the SMM-TF
calculations.
However, in order to draw precise conclusions, a deexcitation treatment that explicitly takes the feeding between known discrete states
of these fragments into account, such as that presented in Ref.\ \cite{ISMMlong},
should be developed and applied to the SMM-TF.

In summary, we have tested, self-consistently, the sensitivity of equilibrium mutifragmentation theories to the density dependence of the
symmetry energy.
Our statistical calculations, which consistently incorporate Skyrme effective interactions,
suggest that the corresponding predicted differences in the isoscaling parameter $\alpha$ are not large enough
to allow one to distinguish between symmetry energies with very different density dependencies.
This observation does not impact the use of $\alpha$ to probe the asymmetry of the emitting system, a use that has been instrumental for
investigations of the density dependent symmetry energy \cite{isoscalingAMD2003,TanisoDiff,TanIsoEOS}.
On the other hand, the isotopic distribution of the fragments produced in the reactions retains some sensitivity to the effective
force employed in the calculations.
We believe that constraining the key SMM-TF inputs, {\it i.e.} the source composition and the excitation energy,
through experimental values of additional observables besides the isoscaling parameters,
will be critical to precision comparisons aimed at constraining the symmetry energy with isotopic distributions.

\begin{acknowledgments}
We would like to acknowledge CNPq, FAPERJ, and the PRONEX program under contract 
No E-26/171.528/2006, for partial financial support.
This work was supported in part by the National Science Foundation under Grant
Nos.\ PHY-0606007 and INT-0228058.
AWS is supported by NSF grant 04-56903.
\end{acknowledgments}

\bibliography{isotf}

\end{document}